\documentclass[aps,prl,reprint]{revtex4-1}
\pdfoutput=1
\usepackage{graphicx}
\usepackage{amsmath,amsfonts,amssymb}
\usepackage{paralist}
\usepackage{caption}
\usepackage{subcaption}
\usepackage{hyperref}
\usepackage[all]{hypcap}
\usepackage{accents}

\newcommand{\antinue}{\ensuremath{\bar{\nu}_e}}
\newcommand{\antinumu}{\ensuremath{\bar{\nu}_\mu}}

\newcommand{\nue}{\ensuremath{\nu_e}}
\newcommand{\numu}{\ensuremath{\nu_\mu}}
\newcommand{\nutau}{\ensuremath{\nu_\tau}}
\newcommand{\aornue}{\ensuremath{\accentset{\left(-\right)}{\nue}}}
\newcommand{\aornumu}{\ensuremath{\accentset{\left(-\right)}{\numu}}}

\newcommand{\ba}{\begin{array}{c}}
\newcommand{\baz}{\begin{array}{cc}}
\newcommand{\bad}{\begin{array}{ccc}}
\newcommand{\bav}{\begin{array}{cccc}}
\newcommand{\baf}{\begin{array}{ccccc}}
\newcommand{\bag}{\begin{array}{cccccc}}
\newcommand{\ea}{\end{array}}

\begin{document}

\title{On the interpretation of IceCube cascade events in terms of the Glashow resonance}

\author{Atri Bhattacharya}
\email{atri@hri.res.in}
\author{Raj Gandhi}
\email{nubarnu@gmail.com}
\affiliation{Harish-Chandra Research Institute, Chhatnag Road, Jhunsi, 211019 Allahabad, India}
\author{Werner Rodejohann}
\email{werner.rodejohann@mpi-hd.mpg.de}
\author{Atsushi Watanabe}
\email{atsushi.watanabe@mpi-hd.mpg.de}
\affiliation{Max-Planck-Institut f\"ur Kernphysik, Saupfercheckweg 1, 69117 Heidelberg, Germany}

\begin{abstract}
The IceCube experiment (\emph{IC}) has recently observed 2 cascade
events with energies between 1 and 10 PeV. 
This energy combined with the fact that no muon-track events are observed 
may be interpreted as a cosmogenic $\antinue$ interacting in IC via the Glashow 
resonance (\emph{GR}) $\bar \nu_e e \to W^-\to$ (hadrons or $\bar
\nu_e e$).
We point out a unique, background-free signature of the GR, a single
isolated muon unaccompanied by any shower activity from the
interaction $\bar \nu_e e \to W^-\to \bar \nu_\mu \mu^-$, and
propose it as a test of this interpretation. We calculate the event
numbers and find that a single such event is expected over about a
three-year period in IC. 
We also show that, if event rates remain at their current levels then, even with the GR, standard cosmogenic fluxes cannot easily explain the observations.
Moreover, if muon-tracks remain conspicuous by their absence, then new physics needs to be invoked.
As example scenarios in conformity with the observations, we calculate event rates for neutrino decay and Lorentz-invariance violation. 
\end{abstract}

\maketitle

\section{\label{sec:intro}Introduction}
Observation of ultra high energy neutrinos would be an important step
towards identifying the origin of cosmic rays and understanding the physics of nature's most powerful accelerators. 
After decades of pioneering work \cite{history}, the
IceCube detector \cite{Halzen:2009tz} is currently
the most sensitive tool to study this problem. Interestingly, two high energy
events, with energy in the range of 1 to 10 PeV, have recently been
reported, with an expected atmospheric background of 0.14 events \cite{talks}.
They are cascade events, implying that they could be caused by \begin{inparaenum}[\itshape a\upshape)]
\item \nue~or \nutau~interacting with nucleons ($N$) via charged current (\emph{CC}) interactions, or,
\item $\nu_\alpha$ ($\alpha = e,\,\mu,\text{ or }\tau$) interacting
with $N$ via neutral current (\emph{NC}) interactions.
\end{inparaenum}
A $\numu N$ CC interaction would have been accompanied by a long
charged muon-track, which is not seen. 
Similarly, a $\nutau N$ CC interaction at these energies would have
been accompanied by a ``double-bang'' \cite{Learned:1994wg}.
Since CC interaction cross sections at these
energies are typically higher than NC cross sections by a factor of
about 2.5, an interpretation via {\itshape a}) is favoured. 

Given the 
present strong constraints on fluxes below $1$ PeV from
IceCube, any future observation in the range $1 - 10$ PeV would likely be due to the Glashow resonance \cite{GR, *Berezinsky:1977sf},
involving the interaction $\bar \nu_e e \to W^-$, where the $W^-$
decays mainly to electrons and hadrons, thus creating cascade events. 
The required resonant energy is $E = m_W^2 /(2 m_e) \simeq 6.3$ PeV,
in the right ballpark vis-a-vis the two IceCube events. 
At energies close to the resonance, the cross section $\antinue e \to
\text{anything}$ is about 350 times the standard CC interaction, $\numu N \to \text{anything}$,
while the cross section for $\antinue e
\to \text{hadrons}$ is about 240 times the standard CC value. 
It is for these reasons that the GR is an important factor in interpreting any observed 
event in the region $1-10$ PeV~\cite{Bhattacharya:2011qu}. 
The GR is also characterised by a unique and essentially background-free signature
(the ``pure muon'' \cite{Bhattacharya:2011qu}), which, 
although suppressed, may be observable
over a three-year period in IC as discussed below. 
These pure, or contained, and isolated muons
 occur when the resonantly produced $W^-$
decays into a muon.
Hence only a contained muon-track (and nothing else) is seen
inside the detector.
Unlike $\nu_\mu N$ CC scattering, there is
no shower activity, and therefore it is easily distinguishable from
``standard'' events. 
We will extract the necessary flux of $\bar\nu_e$ to generate two
events, and apply this number to predict the expected number of ``pure
muons''. 
Since pure muons are essentially background-free, even the small
number of $0.34$ events per year that we find, indicates that in 10 years three or four such events at $E = 6.3$ PeV can be expected and serve both as a straightforward test of
the GR interpretation of IC's two events, as well
as provide a unique physical signature of this important but as yet
undetected Standard Model physics.  

In general, neutrino fluxes from energetic astrophysical sources like active galactic nuclei and gamma-ray bursts are expected to have a significant $\numu + \antinumu$ component which would manifest itself in the form of long muon-tracks in IC. 
So far, IC has not reported any observation of such events at energies around the range spanned by the reported cascade events. 
If indeed, further data gathering and analysis confirms that they are absent or suppressed in the 1--10 PeV range and above it, then this would indicate that the incoming
neutrino flavour composition has a large
$\bar\nu_e/(\nu_\mu + \bar \nu_\mu)$ ratio.
A natural source of mainly electron-antineutrinos are cosmogenic neutrinos \cite{Engel:2001hd,Kotera:2010yn},
which at energies between 1--10 PeV are indeed dominated by $\bar\nu_e$ created in neutron
decay. This interpretation of the two IC events has been put
forward in Ref.~\cite{Barger:2012mz}, and we investigate it further in
this letter both qualitatively and quantitatively. We find that
typical fluxes in the literature have to be scaled up to conform to the observations. 
As at higher energy the cosmogenic neutrinos 
contain significant fraction of muon-neutrinos, we evaluate the implied event
numbers for muon-track events using several typical fluxes for cosmogenic
neutrinos.
These rates lie within the present sensitivity of IC (red curve in Fig.~\ref{fig:non-std}) and should be observable in the near future as more data gathering and analysis is done.

If indeed, muon events continue to remain unobserved at energies between $10^7 - 10^{11}$ GeV, where they are expected to dominate in the case of cosmogenic fluxes, new physics may need to be invoked to account for their absence.
We find that if neutrinos 
decay on their journey to Earth, and if the normal mass ordering is
realized, then the $\aornue$ fluxes
dominate over the \aornumu. In contrast, for an inverted mass
ordering the $\nu_e$ and $\bar \nu_e$ are heavily suppressed. Additionally, we
show that tiny amounts of Lorentz violation could again lead to muon
neutrinos (\numu + \antinumu) being suppressed compared to electron
neutrinos (\nue + \antinue). 

In the following section we recapitulate the physics of the GR and its
consequences for IC. 

\section{\label{sec:GR_std}IceCube and the Glashow Resonance}
We shall assume in this note that the two events
at IC are generated by the Glashow resonance,  
$
\bar \nu_e e \to W^- \to (\mbox{hadrons or } \bar \nu_e e)
$.
It is straightforward to evaluate the flux needed at these energies
that leads to two shower events due to the resonance at IC.
They were observed during the IC designated 2011--2012 86 string period which had a live time of 353.67 days \cite{talks}. Hence, we assume an exposure period of $1 \text{ km}^3\text{-yr}$ in our calculations.
The event rate is \cite{Bhattacharya:2011qu}
\begin{multline}
%\begin{split}
{\rm Rate}(\bar{\nu}_e e \to {\rm hadrons}) \,=\,
2\pi \,\frac{10}{18} N_A \, V_\text{eff} \\
\times\!\int \!\! dE_\nu \int_0^1\!\! dy \,
\frac{d\sigma}{dy}(\bar{\nu}_e e \to {\rm
hadrons})\,\Phi_{\bar{\nu}_e}(E_\nu) \, ,
%\end{split}
\end{multline}
where $N_A = 6.022 \times 10^{23} \,{\rm cm^{-3}}$, $V_{\rm eff} 
\approx 1\text{ km}^3$, and the factor 10/18 takes into account the
number of electrons in water per unit volume.  
The events are integrated over the upper half sphere since up-moving
$\bar \nu_e$ at these energies are attenuated by Earth matter.  
We find
\begin{equation}
\label{eq:flux0}
E^2 \Phi_{\bar{\nu}_e}(E_\nu)\big\vert_{E_\nu = 6.3 \text{ PeV}} = 7.6 \times 10^{-9}
\text{ GeV cm}^{-2}\text{ s}^{-1}\text{ sr}^{-1}.
\end{equation}
This value is in a region that
will be within the reach of IC's detection as it collects more
data. 

Should the two events indeed be a result of the GR, it will be the first observation 
\begin{inparaenum}[\itshape a\upshape)]
\item of the Glashow resonance, and
\item of extra-galactic neutrinos.
\end{inparaenum}
Therefore, it is important to confirm whether the observed events are
indeed due to this process. One way to check this experimentally 
would be to look at the number of ``pure muon'' and ``contained tau
track'' events generated as a result of the related processes 
$ \bar{\nu}_e e \to \bar{\nu}_\mu \mu $ and $ \bar{\nu}_e e \to
\bar{\nu}_\tau \tau $ respectively, which accompany the GR hadronic
interaction and occur when the resonant $W^-$ decays to a muon or
tau instead of hadrons \cite{Bhattacharya:2011qu}. 
Such pure muon events would be clearly distinguishable from the muon-track events coming from the neutrino-nucleon
interactions because, originating in a leptonic interaction, these would not have the characteristic
accompanying hadronic shower. 
The cross section for this process is related to the one to
generate hadrons 
by $\Gamma_{W \to \bar{\nu}_\mu \mu}/\Gamma_{W \to {\rm hadrons}} \approx
0.156$.   
Thus the event rate is small but distinctive in signature.
With the flux $\Phi_{\bar{\nu}_e}$ obtained in Eq.~\eqref{eq:flux0} we
can estimate the pure muon event rate at $E = 6.3$
PeV as $0.34$ per year, leading to a single
event over a span of about three years.  
However, the ``smoking-gun'' nature of these signals, and the fact that the background (from $\numu e \to \nue \mu$ interactions) is negligible \footnote{We estimate the background at these energies to be around $10^{-4}$ events per year.}, makes them a significant tool for GR detection and confirmation in spite of the small rates.

\paragraph*{}
The next step is to consider sources of high energy neutrinos
responsible for the GR events. Diffuse fluxes with a typical power
law will have roughly equal $\nu_\mu$ and $\nu_e$ content when reaching
Earth, and one should expect at energies below PeV a sizable number of
muon-track events, which however, so far, have not been seen. 
It is a useful exercise, therefore, to find neutrino sources circumventing the problem. We will
show in the next section that cosmogenic neutrinos are natural candidates for this.

\section{\label{sec:cosmo}Cosmogenic neutrinos}
In this section we shall assume that the origin of the
electron-antineutrinos is cosmogenic \cite{Engel:2001hd, Kotera:2010yn}, {\it i.e.}, generated by interactions of
cosmic ray protons with the cosmic 
microwave photons, creating a $\Delta$ resonance: 
\begin{equation}
p \gamma \to \Delta^+ \to \pi^+  n \, . 
\end{equation}
At the low energies, $E < 100 $ PeV, the resulting flux is
dominated by the neutron decay, and hence 
\begin{equation}
%\begin{split}
\Phi^0_{\bar{\nu}_e} : \Phi^0_{\bar{\nu}_\mu} :
\Phi^0_{\bar{\nu}_\tau} = 1:0:0 ~\mbox{ and } \Phi^0_{\nu_\alpha} = 0\, ,
%\end{split}
\end{equation}
is the initial flavour composition. Here $\alpha$ represents all flavours of the neutrino 
spectrum.
Lepton mixing will 
modify this flux decomposition, and at Earth we have  
$\Phi_{\nu_\alpha} = P_{\alpha \beta}\, \Phi^0_{\nu_\beta}$, 
where the neutrino mixing probability is $
P_{\alpha \beta} = \sum |U_{\alpha i}|^2\, 
|U_{\beta i}|^2$, with  
$U$ being the lepton mixing matrix.
Inserting the current best-fit values of the oscillation parameters
from Ref.~\cite{Fogli:2012ua}, it turns out that the $(1:0:0)$ flux
composition is transferred to $(0.55:0.25:0.20)$. 
Within $3\sigma$ ranges of the oscillation parameters,
the maximal value of the ratio $\Phi_{\bar{\nu}_e}/\Phi_{\bar{\nu}_\mu}$
is $4.65$.

\paragraph*{}
Detailed studies of the spectrum of cosmogenic neutrinos arriving
at Earth have been made in the literature, and we use the one 
calculated in Ref.~\cite{Engel:2001hd} (ESS) to see if it can explain the
events seen at IC.
We normalise this flux so that the $\bar{\nu}_e$ flux at $\sim10 $ PeV
produces the two events at IC over the appropriate exposure. 
This requires scaling the $\bar{\nu}_e$ flux, and consequently the
fluxes of all flavours, up by a factor of about $300$ from those
obtained in Ref.~\cite{Engel:2001hd}.  
At higher energies, \textit{i.e.}~$10^8 - 10^{10}$ GeV, the decay
chain $\Delta^+ \to \pi^+ n \to e^+ p e \nu_e \bar \nu_e \bar 
\nu_\mu \nu_\mu$ causes a significant bump in the fluxes of the
electron and muon neutrinos, and muon antineutrinos.  
As a result, there will be a significant number of shower and muon-track events observed at IC at these energies. 
Thus, with these fluxes, IC should see about
$15$ muon-track standard CC events per year in the energy range $10^7 -
10^9$ GeV, and about $11$ similar events over energies $10^9 - 10^{11}$
GeV.

For comparison, we have carried out similar event rate
calculations using a different prediction for the
cosmogenic neutrino flux \cite{Kotera:2010yn}. With the {\it
FRII} flux spectra described therein, we find that IC should have
seen $2.2$ shower events, and $3.8$ 
muon-track events per year in the energy range $10^5 - 10^7$ GeV.

Evidently, if the CC muon-tracks are not observed in the near future, then it will clearly be difficult to reconcile the observation of
the two shower events at PeV energies with the lack of
characteristic muon-track events at higher energies using
cosmogenic neutrinos. However, if we allow for the effect of
non-standard physics during the propagation of the neutrinos from 
source to Earth, it might be possible to suppress
the muon neutrino flux while keeping the
$\bar\nu_e$ flux unchanged. If this suppression is
sufficient, it would nicely reconcile the observation of shower
events with the lack of muon-track events. In
the following section we investigate two scenarios which allow this: neutrino decay and Lorentz-invariance violation. 

\section{\label{sec:nonstd}Non-standard physics}
\begin{figure*}
\begin{center}
        \begin{subfigure}{0.48\textwidth}
                \centering
                \includegraphics[width=\textwidth]{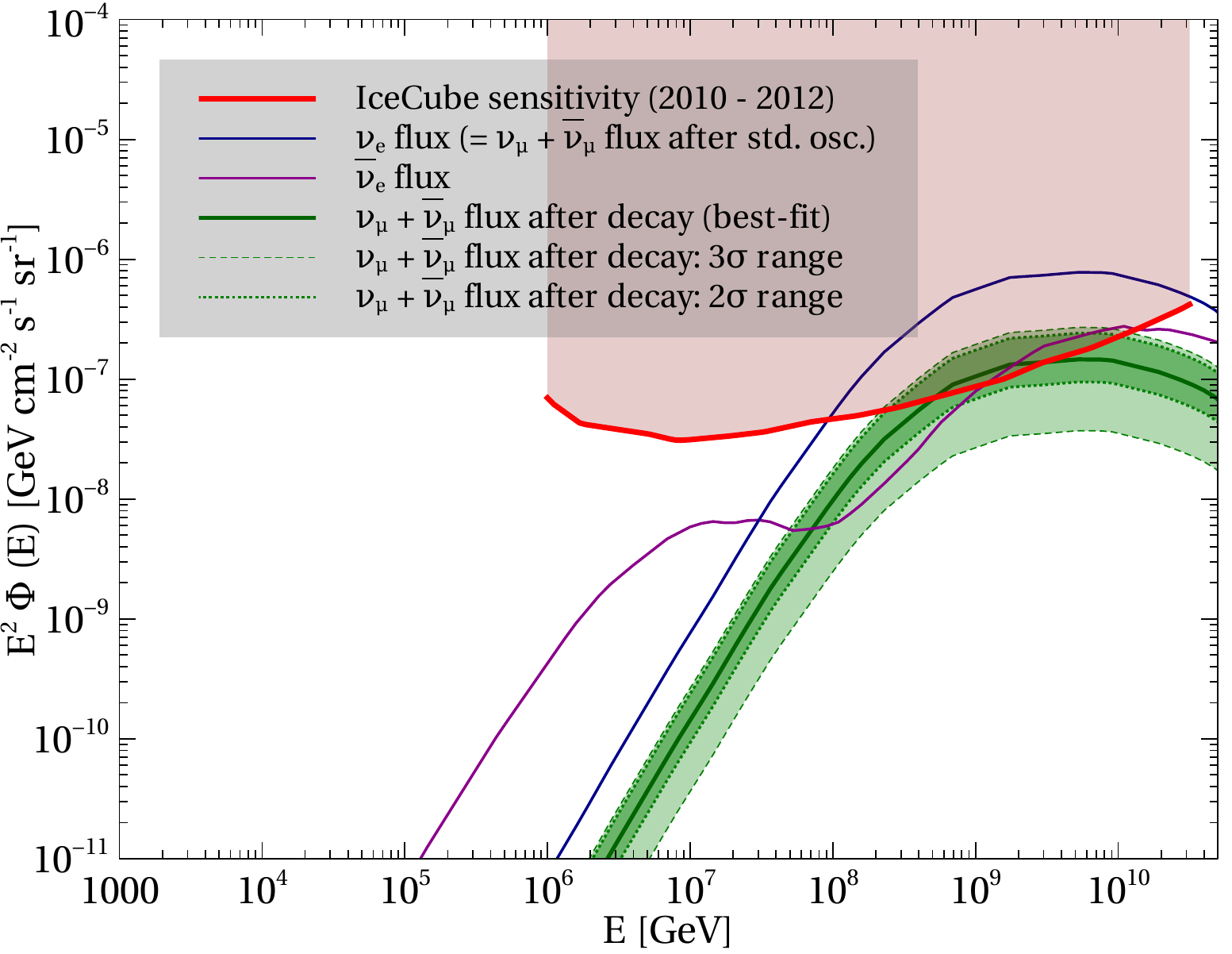}
                \caption{Effect of neutrino decay}
                \label{fig:decay}
        \end{subfigure}%
        \quad %add desired spacing between images, e. g. ~, \quad, \qquad etc. 
          %(or a blank line to force the subfigure onto a new line)
        \begin{subfigure}{0.48\textwidth}
                \centering
                \includegraphics[width=\textwidth]{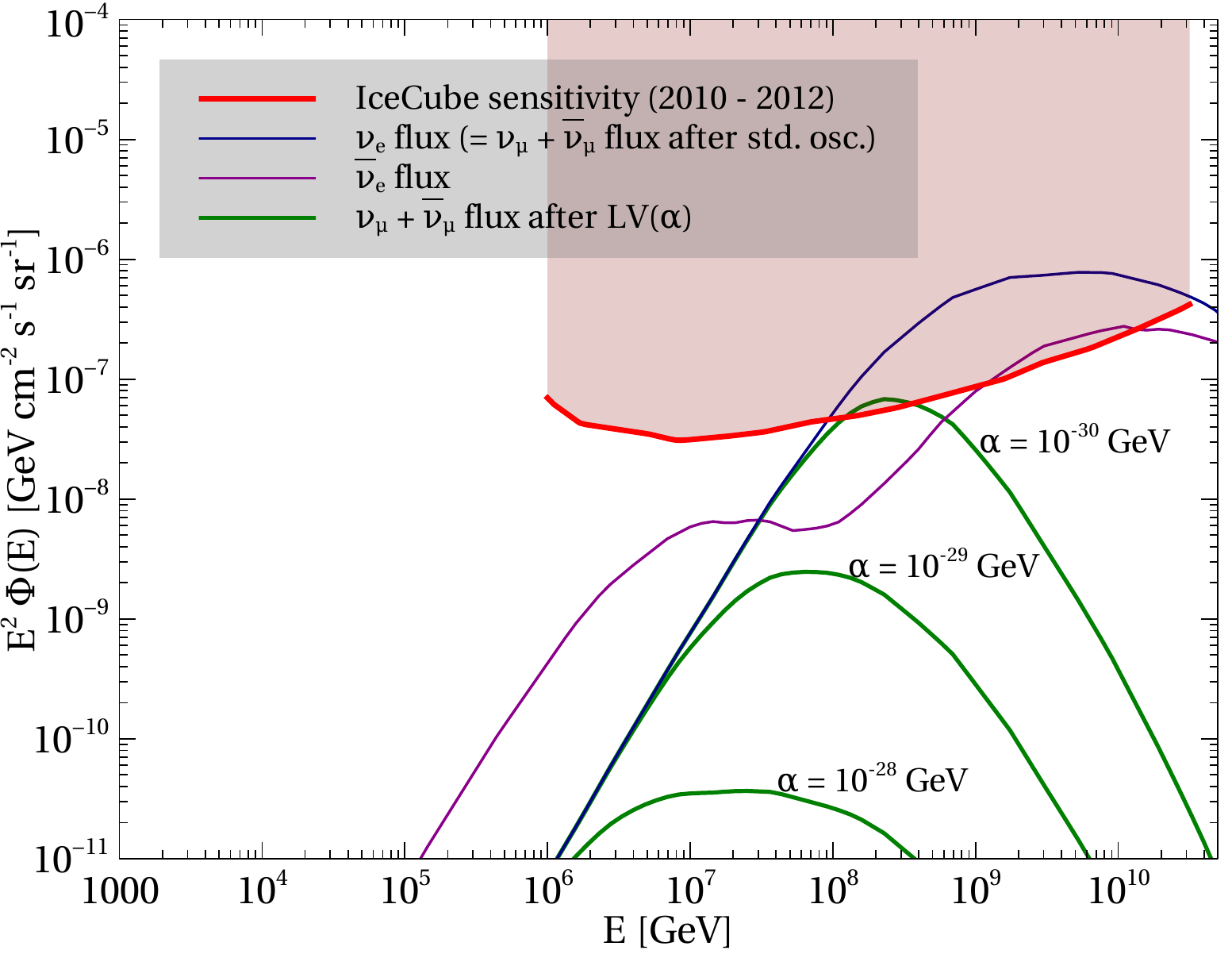}
                \caption{Effect of LV}
                \label{fig:lv}
        \end{subfigure}
\end{center}
\caption{\label{fig:non-std}The effect of non-standard physics, neutrino decay and LV (as
a function of the LV parameter $\alpha$), on the $\nu_\mu$ and
$\bar{\nu}_\mu$ ESS fluxes. The fluxes have all been
appropriately scaled so that the $\bar{\nu}_e$ flux at 6.3 PeV 
explains the two shower events observed at IC. The total $\numu + \antinumu$ flux, when
affected only by standard oscillation, is identical to the $\nue$ flux shown here. For
the decay scenario, the range of variation of the $\nu_\mu$ and
$\bar{\nu}_\mu$ flux as the oscillation parameters vary over the
$2\sigma$ and $3\sigma$ range is also shown.
The projected sensitivity of IC, in the case of a null-event result during the period 2010--2012, is indicated by the red line.} 
\end{figure*}

\paragraph{Neutrino decay.} Assuming that neutrinos produced at the
source decay such that only the lightest
mass state survives, the flux composition
reaching Earth is \cite{Beacom:2002vi}:  
\begin{equation}
\label{eq:ratio}
\left(\Phi_e : \Phi_\mu : \Phi_\tau\right) = \left(|U_{ei}|^2 :
|U_{\mu i}|^2 : |U_{\tau i}|^2\right) ,  
\end{equation}
where $\Phi_\alpha = \Phi_{\nu_\alpha} + \Phi_{\bar{\nu}_\alpha}$ and
either $i = 1$ or $3$, depending on the normal or inverted mass
ordering. In case of the inverted ordering,
since $|U_{e3}|^2 \simeq 0.02 \ll |U_{\mu,\tau 3}|^2 \simeq \frac 12$,
the muon-neutrino flux will dominate and we will run into problems
with non-observation of muon-tracks. Hence we are led to consider the
normal mass ordering \cite{*[{See also }] [{ for an
interpretation of IC's events in terms of neutrino decay.}] Baerwald:2012kc}.
With the present best-fit values of the mixing parameters, one obtains
$\Phi_e/\Phi_\mu = 5.32$. Thus, when the fluxes are scaled so that
the $\bar{\nu}_e$ flux at PeV energies explains the two shower
events seen at IC, the induced suppression of the $\nu_\mu +
\bar{\nu}_\mu$ flux at higher energies leads to about $4.9$ events per year. 
Varying the neutrino mixing parameters over their allowed $3\sigma$
range, we find that the maximum suppression of the muon neutrino flux
with respect to the electron neutrino flux can be $\Phi_e/\Phi_\mu = 20.96$, 
when $\theta_{12} =
30.59^\circ\, , \theta_{23} = 52.96^\circ\, ,\theta_{13} =
10.19^\circ$ with $\delta_{CP} = 180^\circ$. 
With the ESS fluxes scaled as explained in the previous
section, but with the corresponding muon fluxes suppressed by this
factor, the number of muon-track events expected in the energy range
$10^7 - 10^{11}$ GeV reduces to about $1.3$ per year.
Assuming these optimistic values of the mixing parameters, we see that
it is possible to obtain a marginally consistent picture (see Fig.~\ref{fig:decay}): explaining the two shower events at PeV energies via the GR mechanism and sufficiently suppressing any muon-track events.  

\paragraph{Effect of Lorentz-invariance violation (LV).}

Violations of Lorentz-invariance might show up at low energies as a tiny effect in neutrino oscillation \cite{*[{See }] [{ for a recent study of the impact on cosmogenic neutrinos.}] Mattingly:2009jf}.
To simplify the analysis of LV, we study a model where it
affects the oscillation of only $\nu_\mu$ and $\nu_e$.  
The effective Hamiltonian in
the mass basis is modified to 
\begin{equation}
	H_{\mathrm{eff}}(\alpha)=\begin{pmatrix}
			\frac{m_1^2}{2E}	& \alpha 			\\
			\alpha                	& \frac{m_2^2}{2E} 	\\
		\end{pmatrix},
\end{equation}
which gives an oscillation probability of
$
%\begin{equation}
	P\left[ \nu_e \rightarrow \nu_\mu \right] =
\frac{1}{2}\sin^22\theta_{12} \left( 1 - \frac{\alpha^2}{\Omega^2} - \frac{\omega^2}{\Omega^2}\cos \left( 2\Omega L \right) \right)$,
%\end{equation}
where $ \omega=\frac{m^2_2 - m^2_1}{4E} $ and $ \Omega =
\sqrt{\omega^2 + \alpha^2} $. Here $\alpha$ is the LV parameter with
dimensions of energy, and we have only retained the leading $\alpha$
order correction to the probability. 
The effect of LV with various values of $\alpha$ is shown in Fig.~\ref{fig:lv}.
Clearly, even with tiny values of the LV parameter, enough suppression
of the $\nu_\mu$ flux with respect to the $\nu_e$ flux can be achieved
to evade future IC bounds on the muon-track events. For example,
with $\alpha = 10^{-29}$ GeV, about $0.18$ muon-track events are
expected over energies $10^7 - 10^{11}$ GeV per year, which implies
that IC would only be able to observe a single such event over
$5.5$ years of exposure. 
Such a suppression can, therefore, explain the lack of muon-track
events at high energies, while still leaving the $\bar{\nu}_e$
flux at PeV energies large enough to explain the two observed shower
events. 

\paragraph{}
We thus conclude that with non-standard physics it is possible to explain,
simultaneously,
\begin{inparaenum}[\itshape a\upshape)]
\item IC's two shower events at PeV energies (by means of the GR process involving the cosmogenic
$\bar{\nu}_e$), and
\item the non-observation of muon-track events at the
same detector at higher energies.
\end{inparaenum}

\section{\label{sec:concl}Conclusions}
The Glashow resonance is an important Standard Model process which should be
considered in any interpretation of IceCube events seen in the energy range
$10^6-10^7$ GeV. It has a unique and background-free signature, the pure
muon, which can be a smoking gun for this
process. Natural candidates for neutrinos at these (and higher) energies are 
cosmogenic neutrinos. We have shown that standard fluxes,
which have to be scaled by an (unexplained) factor of $300$, do not
explain IC's two events, since charged current generated muon-track events are not seen.
While both the need for scaling the flux up and the paucity of muon-tracks may change as IC collects and analyses new data, we have pointed out that the present observations may,  
however, be understood by invoking the advent of non-standard physics
during propagation. We have demonstrated this with explicit examples of
neutrino decay and Lorentz invariance violation.

\begin{acknowledgments}
We thank Francis Halzen, Aya Ishihara and John Learned for very useful discussions.
WR is supported by the Max Planck Society through the Strategic Innovation Fund and by the DFG in the project RO
2516/4-1, AW by the Young Researcher Overseas Visits Program for Vitalizing Brain Circulation 
Japanese in JSPS.  AB and RG acknowledge the support from the DAE XI Plan Neutrino project.
\end{acknowledgments}

\bibliography{GR_neu}

\end{document}